\RequirePackage[2020-02-02]{latexrelease}
\documentclass[a4paper]{revtex4}
\usepackage{inputenc}
\usepackage[english]{babel}

\usepackage{array,booktabs}% http://ctan.org/pkg/{array,booktabs}
\usepackage{array} 
\usepackage{lipsum}   
\usepackage{calc}
\RequirePackage{graphicx}
\usepackage{color}
\usepackage{amsmath}
\usepackage{float}
\usepackage{calc}
\usepackage{pdflscape}
\usepackage{color}
\usepackage{float}
\usepackage{subfigure}
\usepackage[font=small,labelfont=bf]{caption}
\usepackage{graphicx}
\usepackage{pdflscape}
\usepackage{color}
\usepackage{float}
\usepackage{anyfontsize}
\usepackage[font=small,labelfont=bf]{caption}
\usepackage{graphicx}
\usepackage{amsmath}
\usepackage[nodisplayskipstretch]{setspace}
\renewcommand{\arraystretch}{1.5}
\usepackage{setspace}
\usepackage{tabularx}

\usepackage{float}
\usepackage{color}
\usepackage{amsmath}
\usepackage{float}
\usepackage{calc}
\usepackage{pdflscape}
\usepackage{xcolor}
\usepackage{float}
\usepackage{subfigure}
\usepackage[font=small,labelfont=bf]{caption}
\usepackage{graphicx}
\begin{document}
{\setlength\abovedisplayskip{4pt}}
\title{FCNCs, Proton Stability, $ g_{\mu}-2$ Discrepancy, Neutralino cold Dark Matter in Flipped $SU(5) \times  U(1)_{\chi}$ from $F$ Theory with $ A_{4} $ Symmetry.}
\author{Gayatri Ghosh}
\email{gayatrighdh@gmail.com}
\affiliation{Department of Physics, Gauhati University, Jalukbari, Assam-781015, India}
\affiliation{Department of Physics, Pandit Deendayal Upadhayay Mahavidyalaya, Karimganj, Assam-788720, India}

%\ead{{\color{blue}kalpana@gauhati.ac.in}}
%\ead{{\color{blue}gayatrighsh@gmail.com}}
%\ead{{e-mail: kalpana.bora@gmail.com}}

%\ead{{\color{blue}kalpana@gauhati.ac.in}}
%\ead{{\color{blue}gayatrighsh@gmail.com}}
%\ead{{e-mail: kalpana.bora@gmail.com}}

\begin{abstract}
We predict the low energy signatures of a Flipped $SU(5) \times  U(1)_{\chi}$ effective local model , constructed within the framework of F$-$theory based on $ A_{4} $ symmetry. The Flipped SU(5) model from F Theory in the field of particle physics is prominent due to its ability to construct realistic four$-$dimensional theories from higher$-$dimensional compactifications which necessitates a unified description of the fundamental forces and particles of nature, used for exploring various extensions of the Standard Model. We study Flipped $SU(5) \times  U(1)_{\chi}$ Grand Unified Theories (GUTs) with $ A_{4} $ modular symmetry. In our model due to different modular weights assignments, the fermion mass hierarchy exists with different weighton fields. The constraints on the Dirac neutrino Yukawa matrix allows a good tuning to quark and charged lepton masses and mixings for each weighton field, with the neutrino masses and lepton mixing well determined by the type I seesaw mechanism which occurs at the expense of some tuning which manifests itself in charged lepton flavour violating decays which we explore here. The minimal Flipped $SU(5$) model is supplemented with an extra right$-$handed type and its complex conjugate electron state, $ E_{c} + \bar{E_{c}} $, as well as neutral singlet fields. The $ E_{c} + \bar{E_{c}} $ pair gets masses of the order of TeV which solves the $ g_{\mu}- 2$ discrepancy. The predictions of the model for charged lepton flavour violation decay rate and proton decay could be tested in near future experiments. Also we detect in our model the existence of neutralino, its charge mass and spin via direct and indirect detection. 
\end{abstract}

\maketitle
\section{Introduction}\label{Introduction}
\label{intro}
One promising avenue of research for addressing the enigmatic origin of flavor mixing is exploring the intriguing concept of neutrino oscillations. Neutrinos, being elusive and nearly massless particles, have been found to change their flavors as they propagate through space, which indicates that they possess non$-$trivial mass differences. This phenomenon has opened up exciting possibilities, suggesting the existence of physics beyond the Standard Model. Neutrino oscillations have been a catalyst for theoretical speculation, with proposals ranging from new symmetries and interactions to the introduction of sterile neutrinos that do not participate in the known forces but can mix with the active neutrino states. While this area of research is still an ongoing quest, it offers a potential link between the mysterious flavor mixing phenomenon and deeper underlying principles of particle physics yet to be fully understood. As experimental techniques advance and new data become available, we hope to unlock the secrets of flavor mixing and gain a deeper understanding of the fundamental structure of the universe. 
\par 
Furthermore, the study of modular invariance not only has implications for predicting viable fermion mass textures {\color{red}\cite{FS, FS1}} but also extends its reach to shedding light on the nature of dark matter candidates. In certain string theory constructions, modular symmetries may be intimately connected with the stability of dark matter particles, offering a novel path to explore the mysteries of dark matter within the framework of particle physics and cosmology. As our understanding of these profound symmetries deepens, they hold the potential to unlock new avenues for unifying the fundamental forces and resolving some of the most profound questions in physics.
Modular forms, dependent on the positive integer level $N$ and integer weight $k$, offer a powerful framework for constructing realistic models within the context of the inhomogeneous finite modular group $ \Gamma ^{'}_{N} = \frac{\Gamma}{\Gamma(N)}$ {\color{red}\cite{7}}. Particularly, when the weight $k$ is an even number {\color{red}\cite{FS, FS1}}, these modular forms can be elegantly organized into modular multiplets of $ \Gamma_{N} $, providing a systematic approach to model building in particle physics. Notably, researchers have successfully constructed compelling models based on various levels, including $N = 2, N = 3, N = 4, N = 5, and N = 7$. Remarkably, the versatility of the modular invariance approach also allows for the incorporation of several factorizable {\color{red}\cite{40}} and non-factorizable moduli {\color{red}\cite{50}}, adding more richness to the spectrum of viable models.
\par 
Grand unified theories (GUTs) have been a compelling avenue in particle physics, aiming to unify the three gauge interactions of the Standard Model (SM) group into a more elegant and symmetric structure, often exemplified by $SU(5)$ {\color{red}\cite{52}}. One of the remarkable features of GUTs is the embedding of quark and lepton fields into reduced gauge multiplets, leading to intriguing relations between quark and lepton mass matrices. However, in order to accommodate the observed large lepton mixing angles, the inclusion of family symmetry alongside GUTs becomes well$-$ motivated {\color{red}\cite{53}}. Among the family symmetries, the discrete group $A_{4}$ holds particular significance as it allows for the existence of triplet representations {\color{red}\cite{54}}. Nonetheless, combining $A_{4}$ family symmetry with $SU(5)$ GUTs necessitates delicate vacuum alignment of flavons to break $A_{4}$ , which presents an additional motivation for incorporating modular symmetry. The advent of modular symmetry in the context of GUTs was pioneered in an $(\Gamma_{3}= A_{4}) \times SU(5)$ model{\color{red}\cite{55,56}}, and since then, researchers have pursued the construction of modular $SU(5)$ GUT models based on $(\Gamma_{2}= S_{3}) \times SU(5)$  {\color{red}\cite{57}} and $(\Gamma_{4}= S_{4}) \times SU(5)${\color{red}\cite{58,59,60}}. More recently, investigations into $SO(10)$ GUTs have also benefited from the inclusion of modular symmetry, with studies centered around the $(\Gamma_{3}= A_{4})$ modular symmetry. This interplay between GUTs, family symmetries, and modular symmetry opens up new vistas in theoretical physics, promising a deeper understanding of the fundamental forces and the origins of mass and flavor in the universe.
\par 
In this paper, building upon the motivations discussed earlier, we delve into the investigation of Flipped $SU(5)\times U(1)$ Grand Unified Theories (GUTs) augmented with the elegant $(\Gamma_{3}= A_{4})$ modular symmetry characterized by unique assignments of modular weights. These modular weights play a pivotal role in generating the fermion mass hierarchy, as they are associated with the weighton fields. In such Flipped $SU(5)\times U(1)$ Grand Unified Theories (GUTs) with $(\Gamma_{3}= A_{4})$  modular symmetry, the neutrino sector becomes closely linked to the up$-$type quark mass matrix, leading to an intriguing relationship between the up$-$quarks and the Dirac neutrino mass matrices at the GUT scale, namely $m^{T}_{D} = m_{u}$. However, we seek to evade this constraint by venturing into F$-$theory constructions, where various components of the GUT multiplets may lie on different matter curves. By relaxing this constraint, we can fit the quark and lepton mass matrices as well as quark mixing using a single modulus field $ \tau $. The neutrino masses and lepton mixing, on the other hand, are determined by the type I seesaw mechanism, providing an elegant solution to the intricate puzzle of neutrino mass generation. Moreover, we explore the intriguing possibility of the double seesaw mechanism, facilitated by the existence of extra singlet fields that arise naturally in the framework of Flipped SU(5) GUTs. Through these investigations, we aim to shed light on the rich interplay between modular symmetry, GUTs, and neutrino physics, ultimately contributing to a deeper understanding of the fundamental structure of the universe.
\par
The existence of flavor$-$changing neutral currents {\color{red}\cite{GG}}, transitions between different lepton flavors without changing electric charge, has posed a challenging puzzle in the realm of particle physics. The Standard Model (SM) of particle physics successfully describes the observed quark flavor transitions through the W and Z bosons, known as charged current interactions. However, it predicts the absence of flavor$-$changing neutral currents at tree level, which is in agreement with experimental observations to a remarkable precision.  Some aspects of the prospect of charge lepton flavour violation in a clockwork framework which includes Dirac mass terms and Majorana mas terms for the new clockwork fermions has been extensively studied in {\color{red}\cite{GLHEP}}. The predictions of vanishing $ \theta_{13} $ by Tri Bi Maximal mixing {\color{red}\cite{GNPB}} is owing to its invariance under $ \mu-\tau $ exchange symmetry {\color{red}\cite{GG}}. Small explicit breaking of $ \mu-\tau $ symmetry can generate large Dirac CP violating phase in neutrino oscillations {\color{red}\cite{GC}}. 

Nevertheless, extensions beyond the SM are motivated by several unresolved issues, including the hierarchy problem, dark matter, and neutrino oscillations. The Flipped $SU(5)$ model from F Theory {\color{red}\cite{GF1,GF2,GF3,GF4,GF5,GF6,GF7}}, presents an enticing framework that incorporates grand unification, gauge coupling unification, and addresses the fermion mass hierarchy through additional particle representations. This model exhibits an extended gauge symmetry, $SU(5) \times  U(1)_{\chi}$  {\color{red}\cite{JE,KH}}, with the hypercharge embedded within a larger SU(5) multiplet.
\par 
String model building is a significant endeavor as it enables the transformation of superstring data into testable low-energy predictions. Over the past few decades, researchers have developed numerous models within the framework of heterotic and type IIA string theories, as well as IIB and the geometrically analogous F$-$theory {\color{red}\cite{GC}}. Among these, F$-$ theory stands out as a robust framework for constructing effective field theory models with predictive power, thanks to its provision of convenient tools to implement string rules and principles. One of the key strengths of F-theory lies in its ability to accommodate Grand Unified Theories (GUTs) that are widely accepted to contain the low-energy matter content of the Standard Model. F-theory explores GUTs associated with the exceptional gauge group $E_{8}$, which is connected to the highest geometric singularity of the internal manifold resulting from compactification. Consequently, in F$-$theory compactifications, the geometric properties of these singularities encode crucial information about the effective theory's properties. The observable gauge group is expected to be a subgroup of $E_{8}$, emerging on the world volume of a seven-brane wrapped around a four-manifold surface within the internal six$-$dimensional space. Adding to this geometrical picture are certain seven-brane configurations intersecting over the wrapped surface, representing specific GUTs. The gauge sector of the theory localizes on the world volume of the GUT seven-brane, while matter fields reside on Riemann surfaces known as 'matter curves,' formed by intersections of other seven $-$ branes with the GUT surface. This intricate interplay of geometric features and brane configurations allows for a rich and diverse landscape of string models with the potential to match experimental observations and pave the way for further exploration of fundamental physics.
\par 
In the context of string constructions, including compactifications on del Pezzo surfaces, the Higgs sector's available representations are often limited to fundamental and spinorial ones. However, Grand Unified Theories (GUTs) like standard $SU(5)$ and $SO(10)$ require adjoint or higher representations for breaking the corresponding symmetry. Remarkably, in F$-$theory, the symmetry breaking of a GUT with gauge group $G_{S}$ can be achieved by developing flux along a $U(1)$ factor within $G_{S}$. For instance, the symmetry breaking of $G_{S} = SU(5)$ is accomplished by fluxes turned on along the hypercharge $U(1)_{Y}$ factor. Furthermore, the restriction of fluxes along the matter curves determines the multiplicity of matter content in terms of a few integers associated with those fluxes. The combination of flux and Higgs symmetry breaking mechanisms in the context of the SO(10) gauge group offers promising avenues for achieving symmetry breaking down to the standard model group. The SO(10) model has several attractive features, including the accommodation of all matter fields, including the right-handed neutrino, in a single $SO(10)$ representation, specifically the spinorial 16. In standard GUT approaches where all representations are available, common Higgs symmetry breaking patterns of $SO(10)$ involve intermediate symmetries such as $SU(5)\times U(1)_{\chi}$ and left$-$right symmetric $SU(4)\times SU(2)\times SU(2)$ Pati$-$Salam symmetry. Interestingly, these intermediate symmetries do not require large Higgs representations for their breaking. For example, the well$-$known flipped $SU(5)\times U(1)_{\chi}$ model only requires the 10 + $\bar{10}$ Higgs fields for its breaking, while the Pati$-$Salam symmetry breaking can be realized by a vector$-$like Higgs pair of fields transforming as $(4, 1, 2) + (\bar{4}, 1, 2)$. By exploring the combination of flux$-$induced breaking along $U(1)$ factors and Higgs symmetry breaking mechanisms, the $SO(10)$  gauge group can potentially be reduced to the standard model symmetry in an efficient and elegant manner. This approach holds promise for understanding the particle content and dynamics within the context of grand unified theories beyond the minimal SU(5) construction. Further investigations into the constraints and implications of such combined breaking mechanisms, along with possible restrictions on integer parameters associated with fluxes, will be crucial to advancing our understanding of unified theories and the unification of fundamental forces.
\par 
We will explore the phenomenological consequences and predictions for observable low$-$energy physics within the framework of this $E_{8}-SO(10)$ geometric singularity embedding.
$$SO(10) \supset SU(5)\times U(1)_{\chi} \supset SU(3)_{C} \times SU(2)_{L} \times U(1)_{Y} $$
To shed light on the nature of flavor$-$changing neutral currents, we turn to the $A_{4}$ symmetry, a non$-$ Abelian discrete symmetry group. The $A_{4}$ symmetry has gained considerable attention due to its ability to naturally explain the observed tri-bimaximal pattern of neutrino mixing angles, as experimentally determined by neutrino oscillation experiments. By employing the $A_{4}$ symmetry as a guiding principle, we explore the implications for flavor-changing neutral currents within the Flipped $SU(5$) framework. This research paper aims to investigate the FCNC phenomenon in the Flipped $SU(5)$ model, specifically focusing on the effects of $A_{4}$ symmetry breaking. We shall examine the impact of introducing additional scalar fields and how they interact with the matter fields, leading to the emergence of flavor-changing neutral currents at loop level. Moreover, we will discuss the implications of our findings on experimental observables, such as rare decays and meson mixing processes, which serve as valuable probes for the existence of FCNCs beyond the SM. In summary, this research paper delves into the fascinating interplay between the Flipped $SU(5)$ model and $A_{4}$ symmetry, investigating the emergence of flavor-changing neutral currents. By exploring the effects of $ A_{4} $ symmetry breaking, we aim to shed light on the theoretical implications and experimental consequences of FCNCs within this intriguing framework. Ultimately, this study contributes to our understanding of the fundamental forces governing the universe and offers insights into the nature of particle interactions beyond the Standard Model.
\par 
Upon turning on fluxes along the $U(1)_{\chi}$ factor, the first stage of symmetry breaking takes place, leading to an effective theory that precisely corresponds to the well$-$known flipped $SU(5)$ model. In this context, the fermion content, which originates from the 16 representation of $SO(10)$, is elegantly accommodated in the $10 + \bar{5} + 1$ representations of the flipped $SU(5)$ model. Additionally, the Higgs sector consists of Higgs fields in the $10 + \bar{10}$ representation of SU(5), derived from the $16 + \bar{16}$ of SO(10), and further $5 + \bar{5}$ representations stemming from the $10$ of SO(10). The breaking of the flipped SU(5) model to the standard model gauge symmetry is achieved by developing vacuum expectation values (VEVs) along the pair $10 + \bar{10}$, while the $5 + \bar{5}$ Higgs fields provide the necessary Higgs doublets. Having obtained the final gauge symmetry through the combination of flux and Higgs mechanisms, our investigation centers on analyzing the zero$-$mode spectrum, Yukawa potential, and their fundamental characteristics. Subsequently, we delve into the far-reaching consequences of the model in a diverse array of contemporary processes, including but not limited to neutrino physics, proton decay, leptogenesis, and double beta decay, aiming to discern its viability and potential impact on these crucial aspects of particle physics.
\par 
The stability of the proton, one of the fundamental building blocks of matter, has been a subject of intense investigation in particle physics. While the Standard Model (SM) predicts that the proton is absolutely stable due to the conservation of baryon number, various extensions beyond the SM, including grand unified theories (GUTs), suggest the existence of processes that could lead to its decay.

Among the numerous GUT models proposed, the Flipped SU(5) model extends the gauge symmetry of the SM to $SU(5) \times  U(1)_{\chi}$ and incorporates additional matter and Higgs fields. Notably, it predicts the unification of the three gauge couplings at a high energy scale and provides a natural framework for the observed fermion mass hierarchy. 
In this research paper, we focus on the fascinating connection between proton decay{\color{red}\cite{GPD,GPD1}} and the $A_{4}$ symmetry within the Flipped SU(5) model. The $A_{4}$ symmetry, a non Abelian discrete symmetry group, has garnered significant interest for its ability to explain the observed patterns of lepton flavor mixing angles. By incorporating $A_{4}$ symmetry as a guiding principle, we investigate its impact on proton decay rates and the underlying dynamics responsible for this phenomenon.

The Flipped SU(5) model introduces new gauge bosons, such as the color-triplet X and Y bosons, which can mediate proton decay processes. The $A_{4}$ symmetry plays a crucial role in determining the couplings and interactions of these gauge bosons with quarks and leptons. By imposing the $A_{4}$ symmetry on the model, we explore the implications for proton decay lifetimes, branching ratios, and the predicted experimental signatures.

Probing proton decay is a challenging task that requires dedicated experimental efforts. Current and future experiments, such as Super-Kamiokande, Hyper-Kamiokande, and DUNE, provide powerful tools for testing the predictions of GUT models and probing the validity of the Flipped SU(5) framework. By comparing the theoretical predictions of proton decay rates in the Flipped SU(5) model with the experimental constraints, we aim to shed light on the viability and implications of this model.

In summary, this research paper investigates the phenomenon of proton decay within the context of the Flipped SU(5) model, incorporating the $A_{4}$ symmetry as a guiding principle. By exploring the interplay between proton decay and the underlying dynamics of the Flipped SU(5) framework, we aim to unravel the theoretical implications and potential experimental signatures associated with this intriguing phenomenon. Ultimately, this study contributes to our understanding of the fundamental nature of particle interactions and offers insights into the unification of the fundamental forces of nature beyond the Standard Model.
\par 
The precise measurement of the muon magnetic moment has played a crucial role in scrutinizing the predictions of the Standard Model (SM) of particle physics. However, recent experimental measurements from the Muon g$-$2 Collaboration at Fermilab have unveiled a tantalizing discrepancy between the observed and predicted values. This anomaly, known as the($(g-2)_{\mu}$ discrepancy {\color{red}\cite{GAMM}}, has ignited considerable excitement and has been a topic of intense theoretical and experimental investigation.

One of the most intriguing extensions of the SM that provides a potential solution to the 
$(g-2)_{\mu}$ discrepancy is the Flipped SU(5) model. This model incorporates an extended gauge symmetry of $SU(5) \times  U(1)_{\chi}$ and introduces additional matter and Higgs fields. The Flipped SU(5) model has gained popularity due to its ability to address various fundamental puzzles, such as the gauge coupling unification, fermion mass hierarchy, and neutrino oscillations. 
\par 
In this research paper, we focus on the fascinating interplay between the($(g-2)_{\mu}$ discrepancy and the $A_{4}$ symmetry within the Flipped SU(5) model. The $A_{4}$ symmetry, a non-Abelian discrete symmetry group, has been widely explored for its ability to explain the observed patterns of neutrino mixing angles. By incorporating $A_{4}$ symmetry as a guiding principle, we investigate its role in addressing the($(g-2)_{\mu}$ anomaly and its implications for the underlying dynamics of the Flipped SU(5) framework.

The Flipped SU(5) model introduces new particles and interactions that can contribute to the muon magnetic moment. By considering the impact of $A_{4}$ symmetry on the relevant couplings and particle spectra, we explore the potential enhancements or suppressions of the muon magnetic moment within this framework. Furthermore, we investigate the interplay between the contributions arising from the extended gauge sector and the $A_{4}$ symmetry-breaking sector, providing insights into the potential solutions to the $(g-2)_{\mu}$ discrepancy. 
\par 
Addressing the $(g-2)_{\mu}$ anomaly poses a significant challenge, requiring the development of new theoretical ideas and dedicated experimental efforts. Ongoing experiments, such as the Muon g-2 experiment at Fermilab and future experiments at J PARC and CERN, aim to reduce the uncertainties and further scrutinize the muon magnetic moment. By comparing the predictions of the Flipped SU(5) model with the experimental measurements, we aim to shed light on the viability of this framework in resolving the $(g-2)_{\mu}$ discrepancy. The package micrOMEGAs {\color{red}\cite{64}} embedded in MSSMTools is used here to calculate
the thermally averaged cross section, the DM relic density, and the spin-dependent
(SD) and spin-independent (SI) DM-nucleon cross sections of dark matter $ \bar{\chi}_{1}^{0}$.
\par 
In summary, this research paper investigates the intriguing $(g-2)_{\mu}$ anomaly within the framework of the Flipped SU(5) model, incorporating the $A_{4}$ symmetry as a guiding principle. By exploring the interplay between the $(g-2)_{\mu}$ discrepancy and the underlying dynamics of the Flipped $SU(5)$ framework, we aim to unravel the theoretical implications and potential solutions to this long-standing puzzle. Ultimately, this study contributes to our understanding of the fundamental nature of particle interactions and offers insights into the unification of the fundamental forces of nature beyond the Standard Model.  
\section{The Flipped $ SU(5) $ Framework }
The Flipped $ SU(5) $ Framework model introduces a novel perspective by incorporating flipped representations to achieve a distinct symmetry breaking mechanism within the context of the $SO(10)$ gauge group. Furthermore, the Flipped$ SU(5)$ model, based on the $SU(5)\times U(1)_{\chi}$ gauge symmetry, has gained attention as a potential alternative to the Georgi$-$Glashow $SU(5)$ theory within the framework of superstrings. This model's appeal lies in its unique feature of spontaneously breaking down to the Standard Model (SM) symmetry using only a pair of 10 + 10 Higgs representations, eliminating the need for adjoint Higgs representations. This property aligns well with many string-derived effective models where the adjoint Higgs representation is absent from the massless spectrum. Additionally, the Flipped $SU(5)$ model offers advantages such as a mechanism for doublet-triplet mass splitting in color triplets and the potential realization of an extended seesaw mechanism for neutrino masses by introducing additional neutral singlets. Notably, the hypercharge generator in this model emerges as a combination of the U(1) subgroup within $SU(5)$ and an external abelian factor$U(1)_{\chi}$, highlighting that it is no longer completely embedded within $SU(5)$.

\begin{table}[htb]
\fontsize{5.8 pt}{5.8 pt}\selectfont
\renewcommand{\arraystretch}{1.5}
\begin{center}
\begin{tabular}{|c|c|c|c|c|c|c|c|c|c|c|c|c|c|c|}
\hline 
$ Fields$ & $F_{1} = \lbrace Q_{1},d_{1}^{c},\nu_{1}^{c}\rbrace $  & $F_{2} = \lbrace Q_{2},d_{2}^{c},\nu_{2}^{c}\rbrace$& $ F_{3} = \lbrace Q_{3},d_{3}^{c},\nu_{3}^{c} \rbrace $& $\bar{f}= \lbrace u^{c},L \rbrace$ & $e_{1}^{c} = e^{c}$&$e_{2}^{c} = \mu^{c}$&$e_{3}^{c} = \tau^{c}$&$\nu_{S}$&$H$&$\bar{h}$&$\phi$&$\bar{\phi}$&$\xi$& 
 $Y_{r}^{2\textit{K}}$ \\ 
\hline 
%%\br
 $SU(5) \times U(1)_{\chi}$ & $(10,-\frac{1}{2})$  & $(10,-\frac{1}{2})$ &  $(10,-\frac{1}{2})$ & $(\bar{5},+\frac{3}{2}) $&$(1,-\frac{5}{2})$&$(1,-\frac{5}{2})$&$(1,-\frac{5}{2})$&$(1,0)$&$(10,-\frac{1}{2})$&$(\bar{10},+\frac{1}{2})$&$(5,+1)$&$(\bar{5},-1)$&$(1,0)$&$(1,0)$ \\
\hline
$A_{4}$ &  $1$  & $1$&  $1$& $3 $&$1^{'}$&$1^{''}$&$1$&$3$&$1$&$1$&$1$&$1$&$1$&$r$\\
\hline
 2\textit{k} &  $+3$  & $+2$& $+1$& $-2$&$+6$&$+4$&$+2$&$0$&$0$&$-1$&$0$&$-1$&$-1$& $2\textit{k}$ \\
\hline
\end{tabular}
\end{center}
\caption{Particle Content Of The Model. The transformation properties of leptons, Yukawa couplings, and right-handed neutrino masses within the $SU(5) \times U(1)_{\chi} \times A_{4}$ framework, where 2$ \textit{k} $ represents the modular weight, are essential in understanding the structure of the flavor space. The arrangement of $ \bar{f} $ in the flavor space as $\bar{f} = \lbrace \bar{f_{1}}, \bar{f_{2}}, \bar{f_{3}} \rbrace $ highlights the distinctive role played by fermions and Higgs superfields. Furthermore, the inclusion of the weighton superfield $ \xi $ adds an additional layer of complexity to the overall model.}
\end{table}

The incorporation of the singlet scalar $ \xi $ with non-trivial modular weight plays a crucial role in generating fermion mass hierarchies within the framework, providing a mechanism for understanding the observed spectrum of particle masses and their hierarchies. By examining the impact of the modular symmetry on the flavor structure, we explore the constraints imposed on the Yukawa superpotential. Specifically, for the up quarks, the modular-invariant form of the Yukawa superpotential, as represented in Eq.(28), is given by $ F \bar{f} \tilde{\phi}$. To fully account for the flavor indices and ensure modular invariance, we derive the comprehensive set of superpotential terms that generate quark masses within this framework.

\begin{equation}
W_{u} = \sum_{i =1,2,3} \lambda^{u}_{3i}\tilde{\xi}^{3-i}Y^{2}_{3}F_{i}\bar{f}\tilde{\phi} +\sum_{i =1,2} \lambda^{u}_{2i}\tilde{\xi}^{5-i}Y^{4}_{3}F_{i}\bar{f}\tilde{\phi} +  \lambda^{u}_{11}\tilde{\xi}^{6}Y^{6}_{3,1}F_{i}\bar{f}\tilde{\phi} + \lambda^{u}_{12}\tilde{\xi}^{6}Y^{6}_{3,2}F_{i}\bar{f}\tilde{\phi},
\end{equation}
\begin{equation}
W_{d} = \sum_{i,j =1,2,3} \lambda^{d}_{ij}\tilde{\xi^{8-i-j}}F_{i}F_{j}\phi
\end{equation}
Here $\lambda^{u,d}_{i,j} $ are free parameters and ˜$ \tilde{\xi} = \frac{\xi}{\lambda} $. Terms such as $\tilde{\xi} Y_{3}^{6}F_{2} \bar{f} \tilde{\phi}$ are neglected in the superpotential since they do not lead to significant deviations and generates hierarchical masses for quarks after $ \xi $ acquires VEV. We write explicitly the Yukawa structures for $ Y_{u}, Y_{d} $ as follows:
\begin{equation}
Y_{u} =  {\begin{bmatrix}
\epsilon^{6}Y^{u(6)}_{1} + \epsilon^{4}\lambda^{u}_{21}Y_{1}^{4}+ \epsilon^{2}\lambda^{u}_{31}Y_{1}  & \epsilon^{3}\lambda^{u}_{22}Y_{1}^{4} + \epsilon\lambda^{u}_{32}Y_{1} & \lambda^{u}_{33}Y_{1}\\
\epsilon^{6}Y^{u(6)}_{2} + \epsilon^{4}\lambda^{u}_{21}Y_{2}^{4}+ \epsilon^{2}\lambda^{u}_{31}Y_{2} & \epsilon^{3}\lambda^{u}_{22}Y_{2}^{4} + \epsilon\lambda^{u}_{32}Y_{2} & \lambda^{u}_{33}Y_{2}\\
\epsilon^{6}Y^{u(6)}_{3} + \epsilon^{4}\lambda^{u}_{21}Y_{3}^{4}+ \epsilon^{2}\lambda^{u}_{31}Y_{3}  & \epsilon^{3}\lambda^{u}_{22}Y_{3}^{4} + \epsilon\lambda^{u}_{32}Y_{3} & \lambda^{u}_{33}Y_{3}\\
\end{bmatrix}}^{\dagger},Y_{d} =  {\begin{bmatrix}
\lambda^{d}_{11}\epsilon^{6}& \lambda^{d}_{12}\epsilon^{5} & \lambda^{d}_{13}\epsilon^{4}\\
\lambda^{d}_{13}\epsilon^{4}& \lambda^{d}_{23}\epsilon^{3} & \lambda^{d}_{33}\epsilon^{2}\\
\lambda^{d}_{13}\epsilon^{4}& \lambda^{d}_{23}\epsilon^{3} & \lambda^{d}_{33}\epsilon^{2}\\
\end{bmatrix}}^{*}
\end{equation}

where $Y^{4}_{i}$ for$ i = 1, 2, 3$ are components of modular form $ Y_{3}^{4} $ of weight 4, and $ Y^{u 6}_{i}$ designate the three components of the linear combination of modular forms $\lambda^{u}_{11}Y^{6}_{3_{1}}+ \lambda^{u}_{12}Y^{6}_{3_{2}}$ of modular weight 6. Here $ * $ and $ \dagger $ represent the complex and Hermitian conjugations, respectively. We can diagonalise the $ Y_{u} $ matrices and generate masses for up quarks via the following 
\begin{equation}
Y^{u} = V_{u} diag\lbrace\tilde{Y_{u}}, \tilde{Y_{c}}, \tilde{Y_{t}}\rbrace V_{u}^{' \dagger}
\end{equation}
Similarly, $Y_{d}$ which is a complex and symmetric matix can be diagonalised by the following equation,
\begin{equation}
Y^{d} = V_{d} diag\lbrace\tilde{Y_{d}}, \tilde{Y_{s}}, \tilde{Y_{b}}\rbrace V_{d}^{' \dagger}
\end{equation},
$V_{u,d}$, $V_{u,d}^{\dagger}$ are unitary matrices. The complex $ Y_{u} $ matrix can be further parametrised in terms of $ \epsilon $ which is,
\begin{equation}
Y_{u} = \left[ \begin{bmatrix}
1 & \epsilon \frac{\lambda^{u}_{11}}{\lambda^{u}_{22}} & \epsilon^{2} \frac{\lambda^{u}_{31}}{\lambda^{u}_{33}}\\
-\epsilon \frac{\lambda^{u}_{11}}{\lambda^{u}_{22}} & 1& \epsilon \frac{\lambda^{u}_{32}}{\lambda^{u}_{33}}\\
-\epsilon^{2} \frac{\lambda^{u}_{31}}{\lambda^{u}_{33}} & - \epsilon \frac{\lambda^{u}_{32}}{\lambda^{u}_{33}} & 1\\
\end{bmatrix} \begin{bmatrix}
\epsilon^{6} Y_{1}^{6}& \epsilon^{6} Y_{2}^{6}  & \epsilon^{6} Y_{3}^{6}\\
\epsilon^{3}\lambda^{u}_{22}Y_{1}^{4}& \epsilon^{3}\lambda^{u}_{22}Y_{2}^{4} & \epsilon^{3}\lambda^{u}_{22}Y_{3}^{4}\\
\lambda^{u}_{33}Y_{1}& \lambda^{u}_{33}Y_{2} & \lambda^{u}_{33}Y_{3}\\
\end{bmatrix}\right] ^{*}
\end{equation}
\section{Proton Stability, $g_{\mu}-2$ Discrepancy, Lepton Flavour Violating Decays and Neutralino } 
\begin{equation}
A_{S_{1}} = \sqcap_{i=1}^{3}\left[ \frac{\alpha_{i}(M_{T})}{\alpha_{i}(M_{G})}\right]^{\frac{c_{i}^{1}}{2b_{i}^{3}}}  \times \left[ \frac{\alpha_{i}(M_{SUSY})}{\alpha_{i}(M_{T})}\right]^{\frac{c_{i}^{1}}{2b_{i}^{2}}} \times \left[ \frac{\alpha_{i}(M_{Z})}{\alpha_{i}(M_{SUSY})}\right]^{\frac{c_{i}^{1^{'}}}{2b_{i}^{1}}}
\end{equation}
\begin{equation}
A_{S_{2}} = \sqcap_{i=1}^{3}\left[ \frac{\alpha_{i}(M_{T})}{\alpha_{i}(M_{G})}\right]^{\frac{c_{i}^{2}}{2b_{i}^{3}}}  \times \left[ \frac{\alpha_{i}(M_{SUSY})}{\alpha_{i}(M_{T})}\right]^{\frac{c_{i}^{2}}{2b_{i}^{2}}} \times \left[ \frac{\alpha_{i}(M_{Z})}{\alpha_{i}(M_{SUSY})}\right]^{\frac{c_{i}^{2^{'}}}{2b_{i}^{1}}}
\end{equation}

\begin{equation}
c_{i}^{\left( 1\right) } = \left( -\frac{11}{5}, -3, -\frac{8}{3}\right), c_{i}^{2} = \left( -\frac{23}{15}, -3, -\frac{8}{3}\right) 
\end{equation}

\begin{equation}
c_{i}^{\left( 1^{'}\right) } = \left( -\frac{11}{10}, -\frac{9}{2}, -4 \right), c_{i}^{2^{'}} = \left( -\frac{23}{10}, -\frac{9}{2}, -4 \right) 
\end{equation}

\begin{equation}
b_{i}^{\left( 3\right) } = b_{i}^{2} + n_{T}\left( \frac{2}{5},0,1\right) + n_{D}\left( \frac{3}{5},1,0\right),
\end{equation}

\begin{equation}
b_{i}^{\left( 2\right) } = \left( \frac{33}{5},1,-3\right),
\end{equation}

\begin{equation}
b_{i}^{\left( 1\right) } = \left( \frac{41}{10},-\frac{19}{6},-7\right),
\end{equation}

\begin{equation}
\Gamma_{p\rightarrow \pi^{0}l_{i}^{+}} = k_{\pi}|C_{\pi^{0}l_{i}^{+}}|^{2}\left( A^{2}_{S_{1}}|\frac{1}{M^{2}}+ \left( \frac{m_{u}}{\upsilon_{u}}\right)^{2}\frac{1}{M^{2}_{\bar{\lambda}}}|^{2} + A^{2}_{S_{2}}|\frac{m_{d}m_{l_{i}}}{\upsilon_{d}\upsilon_{d}}\frac{1}{M^{2}_{\bar{\lambda}}}|^{2}\right) 
\end{equation}

\begin{equation}
\Gamma_{p\rightarrow K^{0}l_{i}^{+}} = k_{K}|C_{K^{0}l_{i}^{+}}|^{2}\left( A^{2}_{S_{1}}|\frac{1}{M^{2}}+ \left( \frac{m_{u}}{\upsilon_{u}}\right)^{2}\frac{1}{M^{2}_{\bar{\lambda}}}|^{2} + A^{2}_{S_{2}}|\frac{m_{s}m_{l_{i}}}{\upsilon_{d}\upsilon_{d}}\frac{1}{M^{2}_{\bar{\lambda}}}|^{2}\right) 
\end{equation}

\begin{equation}
k_{\pi} = \frac{m_{p}A_{L}^{2}}{32 \pi}\left( 1-\frac{m^{2}_{\pi}}{m^{2}_{p}}\right)^{2},  k_{K} = \frac{m_{p}A_{L}^{2}}{32 \pi}\left( 1-\frac{m^{2}_{K}}{m^{2}_{p}}\right)^{2},  
\end{equation}

\begin{equation}
C_{\pi^{0}l_{i}} = T_{\pi^{0}l_{i}}\left( U_{L}\right) _{i1}
\end{equation}
\section{The Superpotential and low energy predictions Proton Decay, FCNCs, $ g_{\mu}- 2$ anamoly, Neutralino detection based on modular invariant flipped $ SU(5) $} 
In particular, we assign the matter fields to the fundamental representation $(5, 1)$ and $(1, 5)$ of $SU(5)\times U(1)_{\chi}$, corresponding to the quark and lepton sectors, respectively. Furthermore, the Higgs fields responsible for symmetry breaking are assigned to the representation $(5, 2) + (10, 1) + (1, 2) + (1, 3) + (1, 1)$ in order to achieve the desired Yukawa couplings and generate the masses of the fermions.
\par 
With this notation, the superpotential terms can be expressed in the familiar field theory notation as follows, incorporating the necessary interactions and couplings among the matter and Higgs fields:
\begin{equation}         
\textit{W}= \gamma^{u}_{ij}F_{i}\bar{f_{j}}\bar{h} + \gamma^{d}_{ij}F_{i}F_{j}h + \gamma^{e}_{ij}e^{c}_{i}\bar{f_{j}}h+ \xi\bar{H}F_{i}s\bar{\psi} + \kappa_{mn}\bar{E^{c}_{m}}E^{c}_{n}\sigma+\eta E^{c}_{n}\bar{f_{j}}h_{\chi}+\lambda_{mj}\bar{E^{c}_{m}}e_{j}^{c}\bar{\psi}
\end{equation}
In the above superpotential, the significance of the first three terms in the superpotential and their role in providing Dirac masses to both the charged fermions and the neutrinos is illustrated. It highlights that the up-quark Yukawa coupling, denoted by $ \propto F_{i}\bar{f_{j}}\bar{h}$ , appears at the tree-level, along with the bottom and charged lepton Yukawa couplings. It further explains that in this particular construction, the $U(1)_{Y}$ fluxes are not activated, resulting in no splitting of the$ SU(5)$ representations. Consequently, all three generations of fermions reside on the same matter curve. By utilizing the geometric structure of the theory, it becomes possible to generate fermion mass hierarchies and incorporate the Kobayashi-Maskawa mixing. The tree-level superpotential terms of matter fields are formed at triple intersections, where the matter curves intersect in the extra dimensions. The coefficients of each Yukawa coupling are determined by performing integrals over the overlapping wavefunctions of the corresponding fields at these intersections. 
\begin{equation}
\gamma_{ij} \propto \int \psi_{i} \psi_{j} \phi_{H} dz_{1}\bar{dz_{1}} dz_{2}\bar{dz_{2}}
\end{equation}
The wavefunction of the Higgs field, denoted as $ \phi_{H} $, plays a crucial role in these computations. Detailed calculations of the Yukawa couplings involving matter curves supporting the three generations have demonstrated that hierarchical Yukawa matrices, reminiscent of the Froggatt-Nielsen mechanism, naturally emerge [20–24]. These derived matrices exhibit eigenmasses and mixing patterns that align with experimental observations, providing a promising agreement with empirical data.
\par 
Returning to the superpotential terms, it should be noted that when the Higgs fields $\bar{H}$ and the singlet $\bar{\psi}$ acquire non-vanishing vacuum expectation values (VEVs), a significant consequence arises. Specifically, the last term in the first line of the superpotential generates a mass term that couples the right-handed neutrino with the singlet field. This interaction is crucial in providing a mechanism for neutrino mass generation and the possibility of incorporating neutrino oscillations within the framework of the model.
where 
\begin{equation}
\xi\bar{H\bar{\psi}}F_{i}s = M_{\nu_{i}^{c}s}\nu^{c}s
\end{equation}
 The top Yukawa coupling implies a $3\times3$ Dirac mass for the neutrinos, denoted as $m_{\nu D}= \gamma _{u}^{ij} \left\langle h\right\rangle $, it is essential to consider the additional contribution of a mass term $M_{s}ss$, permitted by the symmetries of the model. Taking these factors into account, the resulting neutrino mass matrix can be expressed as follows, encompassing both the Dirac and additional mass terms:
 \begin{equation}
M_{\nu}  = \frac{1}{\sqrt{3}}\begin{bmatrix}
0 & m_{\nu D}& 0\\
m_{\nu D}^{T} & 0 & M_{\nu_{c}s}^{T}\\
0 & M_{\nu_{c}s} & M_{s}\\
\end{bmatrix}
\end{equation}
In addition to the mentioned terms, it is worth noting that the neutrino mass matrix can potentially incorporate additional non-renormalizable terms, which could further influence its structure and properties. The implications of this mass matrix, including its impact on various lepton flavor and lepton number violating processes, will be thoroughly analyzed in section 6, shedding light on the low-energy consequences of the proposed model.

\begin{center}
\begin{figure*}[htbp]
\centering{
\begin{subfigure}[]{\includegraphics[height=8.8cm,width=8.8cm]{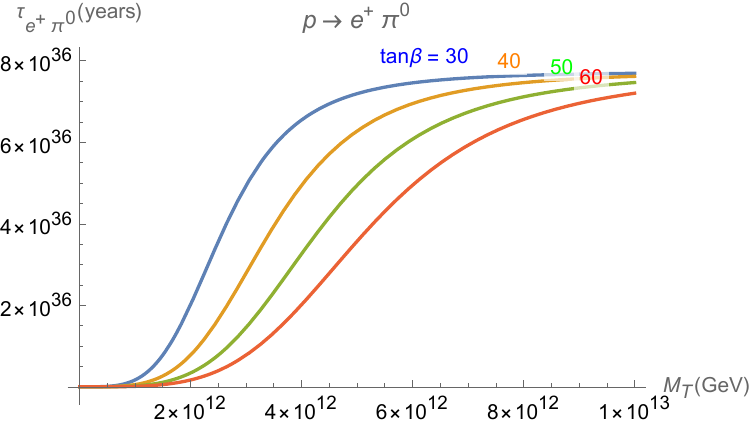}}\end{subfigure}
\begin{subfigure}[]{\includegraphics[height=8.4cm,width=8.8cm]{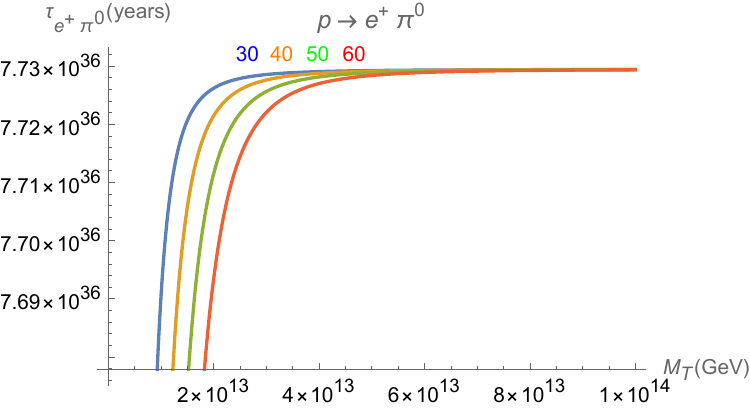}}\end{subfigure}\\
\caption{The results show a clear dependence of the partial-lifetime estimates on the triplet mass $M_{D_{H}^{c}} = \bar{M_{D_{H}^{c}}}= M_{\lambda}$ and the range of $ \tan \beta $ values. By studying the lifetime of the proton through the decay channels $p \rightarrow \pi^{0} + e^{+}$, it has been determined that the allowed range for the triplet mass is bounded at $M_{D_{H}^{c}} = \bar{M_{D_{H}^{c}}}= M_{T} > 10^{11}$ GeV, $M_{G} = 10^{16}$ GeV. Furthermore, it is observed that the asymptotic value of the proton's lifetime is primarily influenced by the masses of the Higgs triplets involved in the decay process.}} 
\label{fig:1}
\end{figure*}
\end{center}

\begin{figure}[h]
\centering
\includegraphics[width=0.6\linewidth]{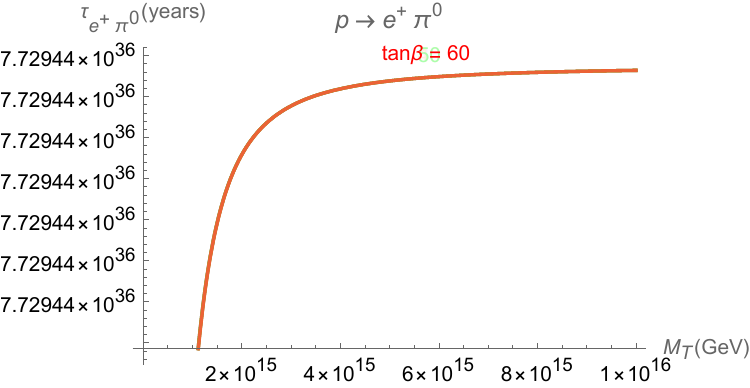}
\caption{In {\color{blue}Fig. 2} The estimates reveal the variation in the partial lifetime of the proton $p \rightarrow \pi^{0} + e^{+}$ for charged lepton decay channels as a function of $M_{D_{H}^{c}} = \bar{M_{D_{H}^{c}}}= M_{T}$, taking into account $ tan \beta = 60 $ values.}
\label{fig:flavour}
\end{figure}

\begin{figure}[h]
\centering
\includegraphics[width=0.6\linewidth]{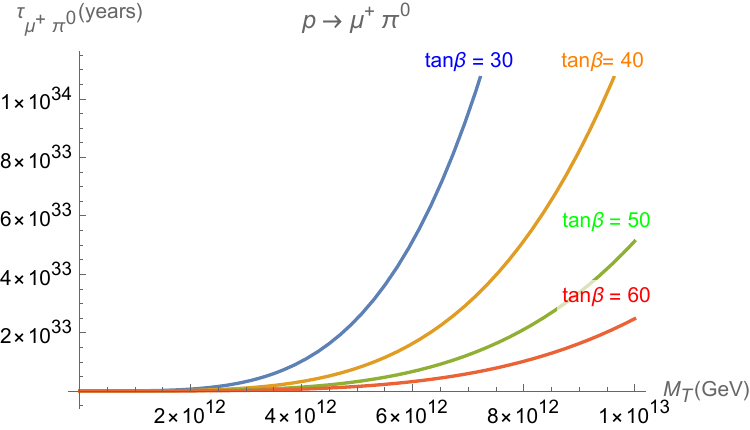}
\caption{The results show a clear dependence of the partial-lifetime estimates on the triplet mass $M_{D_{H}^{c}} = \bar{M_{D_{H}^{c}}}= M_{\lambda}$ and the range of $ \tan \beta $ values. By studying the lifetime of the proton through the decay channels $p \rightarrow \mu^{0} + \pi^{0}$, it has been determined that the allowed range for the triplet mass is bounded at $M_{D_{H}^{c}} = \bar{M_{D_{H}^{c}}}= M_{T} > 6 \times 10^{12}$ GeV, $M_{G} = 10^{16}$ GeV as set by current Super$-$ K bounds. Furthermore, it is observed that the asymptotic value of the proton's lifetime is primarily influenced by the masses of the Higgs triplets involved in the decay process.}
\label{fig:flavour}
\end{figure}

\begin{center}
\begin{figure*}[htbp]
\centering{
\begin{subfigure}[]{\includegraphics[height=8.8cm,width=8.8cm]{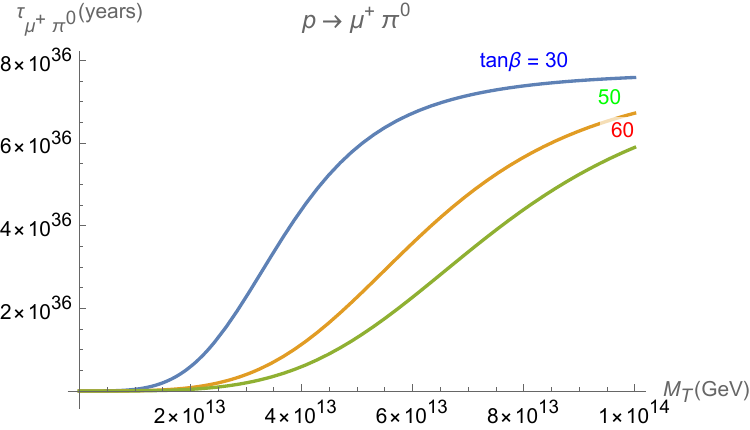}}\end{subfigure}
\begin{subfigure}[]{\includegraphics[height=	8.4cm,width=8.8cm]{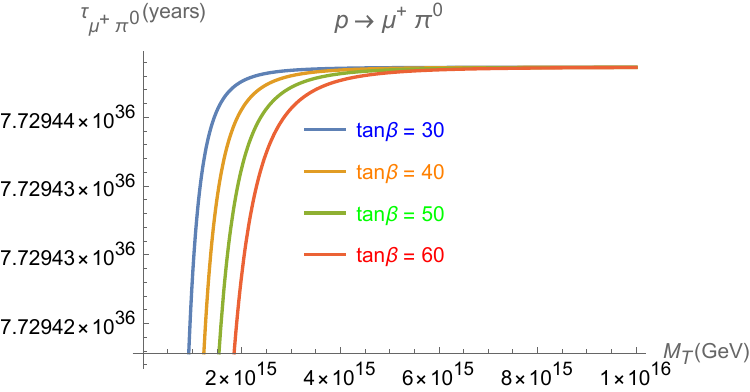}}\end{subfigure}\\
\caption{In {\color{blue}Fig. 4(a), 4(b)} The estimates reveal the variation in the partial lifetime of the proton $p \rightarrow \mu^{+} + \pi^{0}$ for charged lepton decay channels as a function of $M_{D_{H}^{c}} = \bar{M_{D_{H}^{c}}}= M_{T}$, taking into account $ tan \beta = 30, 40, 50, 60 $ values.}}
\label{fig:1}
\end{figure*}
\end{center}

\begin{figure}[h]
\centering
\includegraphics[width=0.6\linewidth]{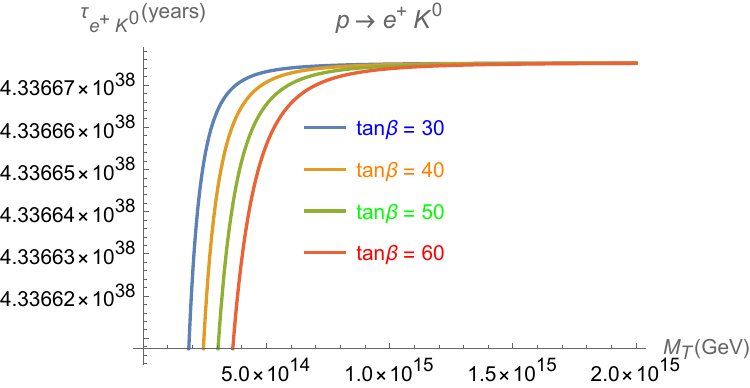}
\caption{In {\color{blue}Fig. 5} the estimates reveal the variation in the partial lifetime of the proton $p \rightarrow e^{+} + K^{0}$ for charged lepton decay channels as a function of $M_{D_{H}^{c}} = \bar{M_{D_{H}^{c}}}= M_{T}$, taking into account $ tan \beta = 30, 40, 50, 60 $ values.}
\label{fig:flavour}
\end{figure}

\begin{center}
\begin{figure*}[htbp]
\centering{
\begin{subfigure}[]{\includegraphics[height=8.8cm,width=8.8cm]{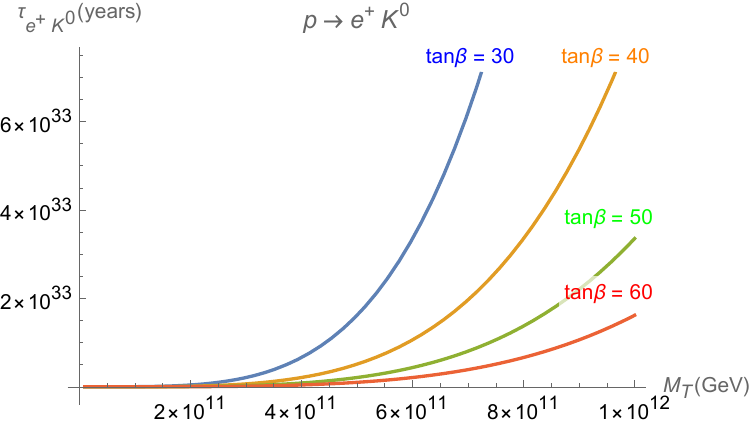}}\end{subfigure}
\begin{subfigure}[]{\includegraphics[height=	8.4cm,width=8.8cm]{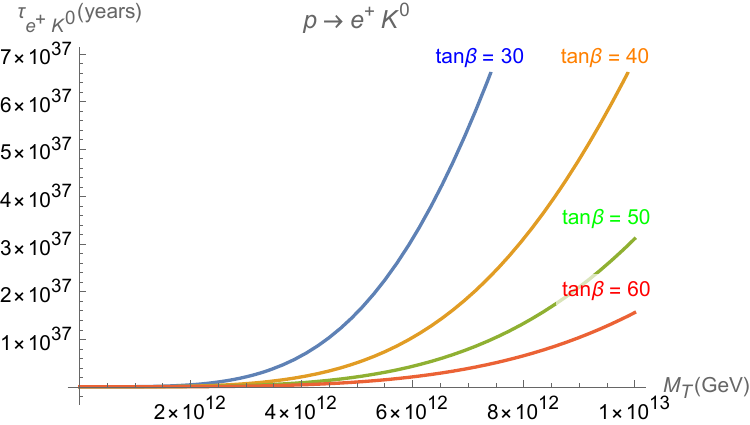}}\end{subfigure}\\

\caption{The results show a clear dependence of the partial-lifetime estimates on the triplet mass $M_{D_{H}^{c}} = \bar{M_{D_{H}^{c}}}= M_{T}$ and the range of $ \tan \beta $ values. By studying the lifetime of the proton through the decay channels $p \rightarrow e^{+} + K^{0}$, it has been determined that the allowed range for the triplet mass is bounded at $M_{D_{H}^{c}} = \bar{M_{D_{H}^{c}}}= M_{T} >  10^{11}$ GeV, $M_{G} = 10^{16}$ GeV as set by current Super$-$ K bounds. Furthermore, it is observed that the asymptotic value of the proton's lifetime is primarily influenced by the masses of the Higgs triplets involved in the decay process.}}
\label{fig:1}
\end{figure*}
\end{center}

\begin{center}
\begin{figure*}[htbp]
\centering{
\begin{subfigure}[]{\includegraphics[height=8.8cm,width=8.8cm]{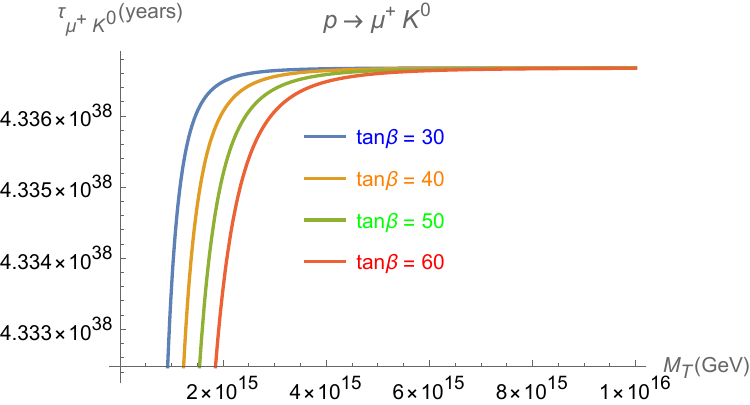}}\end{subfigure}
\begin{subfigure}[]{\includegraphics[height=8.4cm,width=8.8cm]{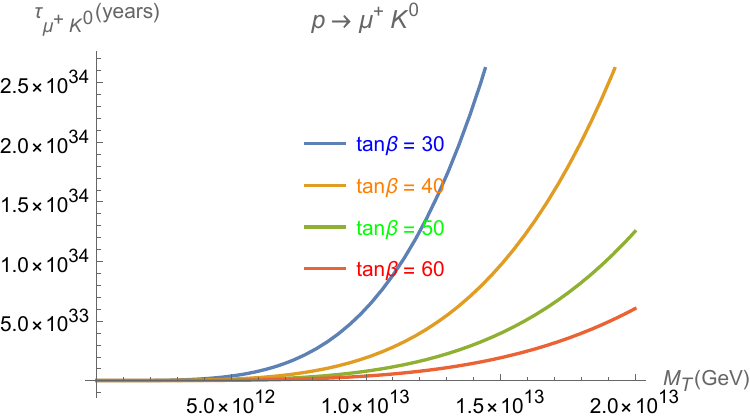}}\end{subfigure}\\
\caption{In {\color{blue}Fig. 7(a), 7(b)} the results show a clear dependence of the partial-lifetime estimates on the triplet mass $M_{D_{H}^{c}} = \bar{M_{D_{H}^{c}}}= M_{T}$ and the range of $ \tan \beta $ values. By studying the lifetime of the proton through the decay channels $p \rightarrow \mu^{+} + K^{0}$, it has been determined that the allowed range for the triplet mass is bounded at $M_{D_{H}^{c}} = \bar{M_{D_{H}^{c}}}= M_{T} >  10^{11}$ GeV, $M_{G} = 6 \times 10^{16}$ GeV as set by current Super$-$ K bounds. Furthermore, it is observed that the asymptotic value of the proton's lifetime is primarily influenced by the masses of the Higgs triplets involved in the decay process.}} 
\label{fig:1}
\end{figure*}
\end{center}

\begin{figure}[h]
\centering
\includegraphics[width=0.6\linewidth]{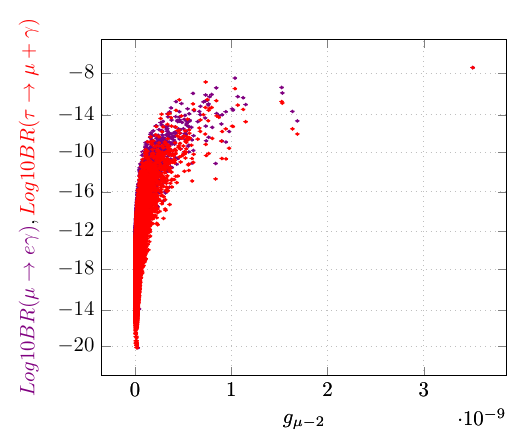}
\caption{Different values of $ g_{\mu}-2 $ discrepancy as a function of $ Log 10 BR(\mu \rightarrow e \gamma, \tau \rightarrow \mu \gamma) $ for calculated values of Dirac Neutrino Yukawa couplings in this work.}
\label{fig:flavour}
\end{figure}

\begin{figure}[h]
\centering
\includegraphics[width=0.6\linewidth]{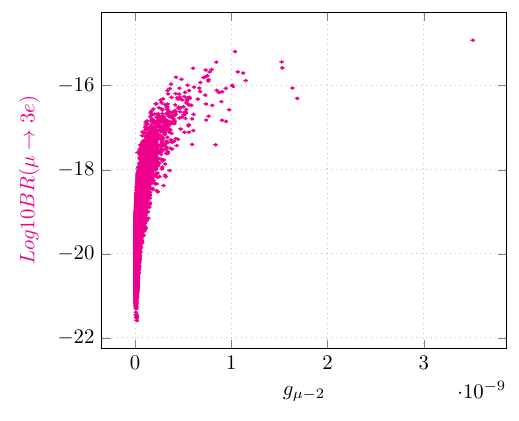}
\caption{Different values of $ g_{\mu}-2 $ discrepancy as a function of $ Log 10 BR(\mu \rightarrow 3e \gamma) $ for calculated values of Dirac Neutrino Yukawa couplings in this work.}
\label{fig:flavour}
\end{figure}

\begin{figure}[h]
\centering
\includegraphics[width=0.6\linewidth]{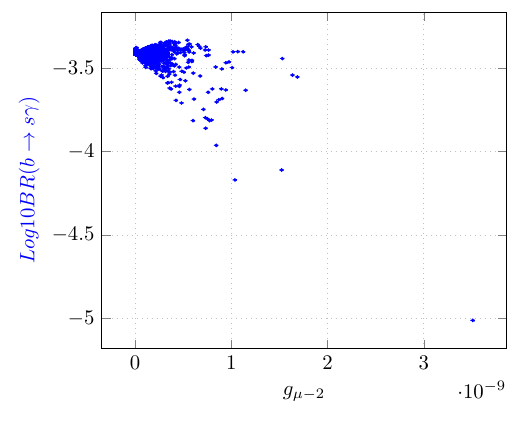}
\caption{Different values of $ g_{\mu}-2 $ discrepancy as a function of $ Log 10 BR(b \rightarrow s \gamma) $ for calculated values of Dirac Neutrino Yukawa couplings in this work.}
\label{fig:flavour.}
\label{fig:flavour}
\end{figure}

\begin{figure}[h]
\centering
\includegraphics[width=0.6\linewidth]{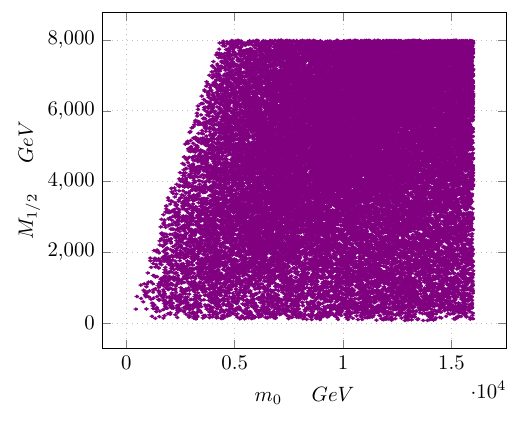}
\caption{Common Values of Universal scalar mass $ m_{0}$ and common univeral gaugino masses $ M_{\frac{1}{2}} $ as constrained by latest updated values on MEG Experiment.}
\label{fig:flavour}
\end{figure}

\begin{figure}[h]
\centering
\includegraphics[width=0.6\linewidth]{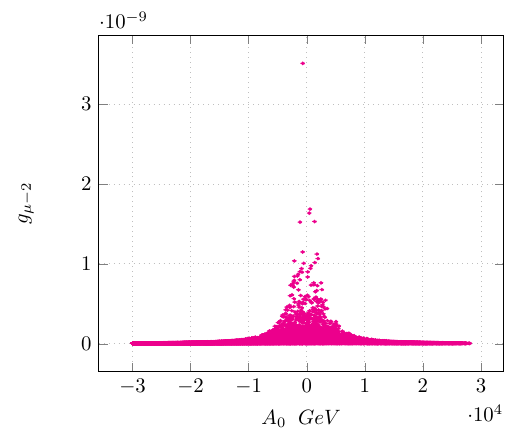}
\caption{{Different values of $ g_{\mu}-2 $ discrepancy as a function of $A_{0} $ for calculated values of Dirac Neutrino Yukawa couplings in this work.}
\label{fig:flavour.}}
\label{fig:flavour}
\end{figure}

\begin{figure}[h]
\centering
\includegraphics[width=0.6\linewidth]{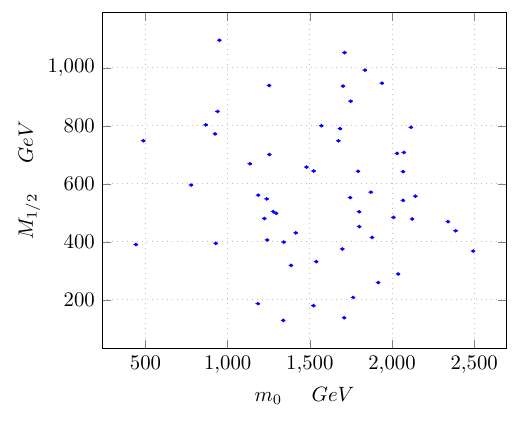}
\caption{Common Values of Universal scalar mass $ m_{0}$ and common univeral gaugino masses $ M_{\frac{1}{2}} $ as constrained by latest updated values on $ b\rightarrow s \gamma $}
\label{fig:flavour}
\end{figure}
From computed universal gaugino masses and common scalar masses at GUT scale we find with the help of software micromegas that dark matter candidate is neutralino with $spin=\frac{1}{2}$  $mass=1.96 \times 10^{3}$ GeV. The calculated masses  Of  HIGGS and other SUSY particles are as follows:
Higgs masses and widths are as follows:
  $$  \text{SM Higg}s =   125.44 \text{ GeV with width} 5.48\times 10^{-03} $$
   $$ \text{CP odd Higgs} = 200.00 \text{ GeV with width} 3.88\times 10^{+02} $$
   $$ \text{Charged Higgs} = 250.43 \text{ GeV with width} 3.88\times 10^{+02} $$

Masses of odd sector Particles in GeV :
$$\sim o1   : MNE1    = 1960.807 , \sim 1+ : MC1     = 3536.172 , \sim o2   : MNE2    = 3536.460$$ 
$$\sim o3   : MNE3    = 3541.697 , \sim ne : MSne    = 3999.510 , \simeq nm : MSnm    = 3999.510$$ 
$$\sim eL   : MSeL    = 4000.241 , mL      : MSmL    = 4000.241 , \sim 2+   : MC2     = 4110.524$$ 
$$\sim o4   : MNE4    = 4110.525 , \sim nl : MSnl    = 5999.673 , \sim l1   : MSl1    = 5999.943 $$
$$\sim g    : MSG     = 6910.993 , \sim eR : MSeR    = 7000.142 , \sim mR   : MSmR    = 7000.142 $$
$$\sim dR   : MSdR    = 7277.285 , \sim sR   : MSsR    = 7277.285 , \sim l2   : MSl2    = 9000.256 $$
$$\sim b1  : MSb1    = 9209.131 , \sim uL    : MSuL    = 10189.006 , \sim cL      : MScL    = 10189.006 $$
$$ \sim dL      : MSdL    = 10189.293 ,  \sim sL      : MSsL    = 10189.293 , \sim t1      : MSt1    = 11885.293 $$
$$\sim b2      : MSb2    = 12165.978 , \sim uR      : MSuR    = 13159.630 , \sim cR      : MScR    = 13159.630 $$
$$\sim t2      : MSt2    = 14474.603 , $$

The calculated relic density comes out to be 
$$  \Omega _{h}=2.78 \times 10^{2}$$.
 Decay Channels contributing to $ \frac{1}{\Omega} $ with more than 1 $\%$ are.
  Relative contributions in percentage are displayed
   $$21 \% \sim o1   \sim o1 \longrightarrow m M $$
   $$21 \% \sim o1   \sim o1 \longrightarrow e E $$
   $$13\% \sim o1   \sim o1 \longrightarrow W^{+} W^{-} $$ 
   $$8\% \sim o1   \sim o1 \longrightarrow l L $$ 
    $$6\% \sim o1   \sim o1 \longrightarrow t T $$ 
   $$6\% \sim o1   \sim o1 \longrightarrow z h $$
    $$5\% \sim o1   \sim o1 \longrightarrow ne NE $$ 
    $$5\% \sim o1   \sim o1 \longrightarrow nm NM $$ 
    $$4\% \sim o1   \sim o1 \longrightarrow h h $$
    $$4\% \sim o1   \sim o1 \longrightarrow Z$$
    $$ 2\%  \sim o1 \sim o1 \rightarrow nl Nl$$ 
    $$ 1\% \sim o1 \sim o1 \rightarrow b B$$ 

The  annihilation cross section of neutralino in $cm^{3}$/s  via indirect detection of $\sim o1$, neutralino  via  various decay channels of neutralino have been computed in our model. The contribution of various decay channels are as follows:

 $$ \sim o1, \sim o1 \rightarrow  t T \text{of annihilation cross section} 8.24\times 10^{-01} cm^{3}/s $$
 $$ \sim o1,\sim o1 \rightarrow b B  \text{of annihilation cross section}  1.66 \times 10^ -{01} cm^{3}/s$$
 $$ \sim o1,\sim o1 \rightarrow l L \text{of annihilation cross section}        8.32\times 10^{-03} cm^{3}/s $$
 $$ \sim o1,\sim o1 \rightarrow A W^{+} W^{-} \text{of annihilation cross section}     3.80\times 10^{-04} cm^{3}/s $$
$$  \sim o1,\sim o1 \rightarrow A m M  \text{of annihilation cross section}     3.15\times 10^{-04} cm^{3}/s $$
$$  \sim o1,\sim o1 \rightarrow A e E \text{of annihilation cross section}      3.15\times 10^{-04} cm^{3}/s $$
$$  \sim o1,\sim o1 \rightarrow W^{+} W^{-} \text{of annihilation cross section}       1.76\times 10^{-04} cm^{3}/s$$
$$  \sim o1,\sim o1 \rightarrow Z Z  \text{of annihilation cross section}       1.36\times 10^{-04 } cm^{3}/s$$
$$  \sim o1,\sim o1 \rightarrow Z h  \text{of annihilation cross section}       1.03\times 10^{-04}  cm^{3}/s$$

The Photon flux  for angle of sight 0.10 radian and spherical region described by cone with angle 0.0349 radian are found to be 7.08 $\times 10 ^{-24}\left[ cm^{2} s GeV\right] ^{-1}$ for E=980.4 GeV. The Positron flux is found to be 9.81 $\times 10 ^{-22} \left [cm^{2} sr s GeV \right ]^{-1}$ for E=980.4 GeV and the antiproton flux is found to be 3.86E $\times 10 ^{-21} \left [cm^{2} sr s GeV\right ]^{-1}$ for E=980.4 GeV. 

\begin{table}[htb]
\renewcommand{\arraystretch}{1.5}
\begin{center}
\begin{tabular}{|c|c|c|}
\hline 
 Branching Ratio &  Partial width & Decay  Channel\\ 
\hline 
%%\br
  $4.608\times 10^{-01}$ & $2.523\times 10^{-03}$\hspace{0.1cm} GeV &$h\rightarrow b,B$\\
\hline
  $3.452\times 10^{-01}$ & $1.890\times 10^{-03}$\hspace{0.1cm} GeV &$h\rightarrow W^{+}, W^{-}$\\
\hline
   $6.917\times 10^{-02}$ & $3.788\times 10^{-04}$\hspace{0.1cm} GeV &$h\rightarrow G, G$\\

\hline
$5.08\times 10^{-02}$ & $2.784 \times 10^{-04}$\hspace{0.1cm} GeV &$h\rightarrow l, L$\\ 
\hline
$4.583\times 10^{-02}$ & $2.510\times 10^{-04}$\hspace{0.1cm} GeV &$h\rightarrow Z,Z $\\ 
\hline
$2.557\times 10^{-02}$ & $1.400\times 10^{-04}$\hspace{0.1cm} GeV &$h\rightarrow c,C$\\
\hline
  $1.209\times 10^{-04}$ & $6.619\times 10^{-07}$\hspace{0.1cm} GeV &$h\rightarrow d,D$\\
\hline
  $1.209\times 10^{-04}$ & $6.619\times 10^{-07}$\hspace{0.1cm} GeV &$h\rightarrow s, S$\\
\hline
   $1.208\times 10^{-04}$ & $6.616\times 10^{-07}$\hspace{0.1cm} GeV &$h\rightarrow u, U$\\

\hline
$2.257\times 10^{-03}$ & $1.236\times 10^{-05}$\hspace{0.1cm} GeV &$h\rightarrow A, A$\\ 
\hline
\end{tabular}
\end{center}
\caption{Branching Ratio, Partial Width and Decay channel of SM Higgs of total width $ 5.4 \times10^{-03} $ .}
\end{table}

\begin{table}[htb]
\renewcommand{\arraystretch}{1.5}
\begin{center}
\begin{tabular}{|c|c|c|}
\hline 
 Branching Ratio &  Partial width & Decay  Channel\\ 
\hline 
%%\br
  $7.030\times 10^{-01}$ & $2.728\times 10^{+02}$\hspace{0.1cm} GeV &$H_{3}\rightarrow b,B$\\
\hline
  $1.208\times 10^{-01}$ & $4.688\times 10^{+01}$\hspace{0.1cm} GeV &$H_{3}\rightarrow l, L$\\
\hline
   $4.960\times 10^{-02}$ & $1.925\times 10^{+01}$\hspace{0.1cm} GeV &$H_{3}\rightarrow \sim 1+                                                  , \sim 2-$\\
\hline
$3.566\times 10^{-02}$ & $1.384\times 10^{+01}$\hspace{0.1cm} GeV & $H_{3}\rightarrow \sim o2, \sim 04$\\
\hline
$1.402E^{-02}$ & $5.443\times 10^{+00}$\hspace{0.1cm} GeV & $H_{3}\rightarrow \sim o3, \sim 04$\\ 
\hline
 $1.353E^{-02}$ & $5.2513\times 10^{+00}$\hspace{0.1cm} GeV & $H_{3}\rightarrow \sim o1, \sim o2$\\ 
\hline
$8.9112E^{-03}$ & $3.459\times 10^{+00}$\hspace{0.1cm} GeV & $H_{3}\rightarrow \sim o1, \sim o3$\\ 
\hline
 $1.713E^{-03}$ & $6.650\times 10^{-01}$\hspace{0.1cm} GeV & $H_{3}\rightarrow \sim 1+, \sim 1-$\\ 
\hline
 $1.045E^{-03}$ & $4.057\times 10^{-01}$\hspace{0.1cm} GeV & $H_{3}\rightarrow \sim 2+, \sim 2-$\\ 
\hline  
 $5.890E^{-04}$ & $2.286\times 10^{-01}$\hspace{0.1cm} GeV & $H_{3}\rightarrow \sim o2, \sim o2$\\ 
\hline
 $5.245E^{-04}$ & $2.036\times 10^{-01}$\hspace{0.1cm} GeV & $H_{3}\rightarrow \sim o4, \sim o4$\\
 \hline
  $5.890E^{-04}$ & $2.286\times 10^{-01}$\hspace{0.1cm} GeV & $H_{3}\rightarrow \sim o2, \sim o2$\\ 
 \hline
  $5.245E^{-04}$ & $2.036\times 10^{-01}$\hspace{0.1cm} GeV & $H_{3}\rightarrow \sim o4, \sim o4$\\
   \hline
  $3.843E^{-04}$ & $1.492\times 10^{-01}$\hspace{0.1cm} GeV & $H_{3}\rightarrow t, T$\\  
   \hline
  $2.604E^{-04}$ & $1.011\times 10^{-01}$\hspace{0.1cm} GeV & $H_{3}\rightarrow D, d$\\ 
   \hline
  $2.604E^{-04}$ & $1.011\times 10^{-01}$\hspace{0.1cm} GeV & $H_{3}\rightarrow S, s$\\ 
   \hline
  $1.056E^{-04}$ & $4.098\times 10^{-02}$\hspace{0.1cm} GeV & $H_{3}\rightarrow \sim o2, \sim o3$\\ 
   \hline
  $2.243E^{-05}$ & $8.708\times 10^{-03}$\hspace{0.1cm} GeV & $H_{3}\rightarrow \sim o1, \sim o4$\\
   \hline
  $2.060E^{-05}$ & $7.996\times 10^{-03}$\hspace{0.1cm} GeV & $H_{3}\rightarrow \sim o1, \sim o1$\\ 
   \hline
  $7.437E^{-06}$ & $2.887\times 10^{-03}$\hspace{0.1cm} GeV & $H_{3}\rightarrow \sim o3, \sim o3$\\
  \hline  
  $4.017E^{-06}$ & $1.559\times 10^{-03}$\hspace{0.1cm} GeV & $H_{3}\rightarrow G,G$\\ 
   \hline
  $2.820E^{-08}$ & $1.095\times 10^{-05}$\hspace{0.1cm} GeV & $H_{3}\rightarrow Z, h$\\
   \hline
  $6.880E^{-09}$ & $2.670\times 10^{-06}$\hspace{0.1cm} GeV & $H_{3}\rightarrow A,A$\\  
   \hline
  $5.9360E^{-09}$ & $2.304\times 10^{-06}$\hspace{0.1cm} GeV & $H_{3}\rightarrow c, C$\\ 
   \hline
  $5.221E^{-11}$ & $2.027\times 10^{-08}$\hspace{0.1cm} GeV & $H_{3}\rightarrow u, U$\\
  \hline 
\end{tabular}
\end{center}
\caption{Branching Ratio, Partial Width and Decay channel of CP odd Higgs of total width $ 3.88 \times10^{+02} $ .}
\end{table}

For $E> 1 $ GeV the neutrino fluxes from Sun calculated in our model are found to be 629 $ \frac{1}{\text{Year}Km^{2}}$. For $E> 1 $ GeV the calculated upward muon fluxes from Sun is  calculated in our model are found to be $6.24 \times 10^{-06} \frac{1}{\text{Year}Km^{2}}$, and for $E> 1 $ GeV the contained muon fluxes from Sun calculated in our model is found to be $2.09 \times 10^{-05}  \frac{1}{\text{Year}Km^{2}}$.

\begin{table}[htb]
\renewcommand{\arraystretch}{1.5}
\begin{center}
\begin{tabular}{|c|c|c|}
\hline 
 Branching Ratio &  Partial width & Decay  Channel\\ 
\hline 
%%\br
  $7.029\times 10^{-01}$ & $2.728\times 10^{+02}$\hspace{0.1cm} GeV &$H \rightarrow b,B$\\
\hline
  $1.208\times 10^{-01}$ & $4.688\times 10^{+01}$\hspace{0.1cm} GeV &$H \rightarrow l, L$\\
\hline
   $5.043\times 10^{-02}$ & $1.958\times 10^{+01}$\hspace{0.1cm} GeV &$H \rightarrow \sim 1-                                                , \sim 2+$\\
\hline
$3.655\times 10^{-02}$ & $1.419\times 10^{+01}$\hspace{0.1cm} GeV & $H \rightarrow \sim o3, \sim o4$\\
\hline
$1.3712E^{-02}$ & $5.320\times 10^{+00}$\hspace{0.1cm} GeV & $H_{3}\rightarrow \sim o2, \sim 04$\\ 
\hline
 $1.351E^{-02}$ & $5.2433\times 10^{+00}$\hspace{0.1cm} GeV & $H \rightarrow \sim o1, \sim o3$\\ 
\hline
$8.933E^{-03}$ & $3.467\times 10^{+00}$\hspace{0.1cm} GeV & $H \rightarrow \sim o1, \sim o2$\\ 
\hline
 $7.987E^{-04}$ & $3.100\times 10^{-01}$\hspace{0.1cm} GeV & $H\rightarrow \sim 1+, \sim 1-$\\ 
\hline
 $3.843E^{-04}$ & $1.491\times 10^{-01}$\hspace{0.1cm} GeV & $H \rightarrow \sim t, T$\\ 
\hline  
 $3.087E^{-04}$ & $1.198\times 10^{-01}$\hspace{0.1cm} GeV & $H \rightarrow \sim 2+, \sim 2-$\\ 
\hline
 $2.716E^{-04}$ & $1.054\times 10^{-01}$\hspace{0.1cm} GeV & $H \rightarrow \sim o2, \sim o2$\\
 \hline
  $2.604E^{-04}$ & $1.011\times 10^{-01}$\hspace{0.1cm} GeV & $H \rightarrow d, D$\\ 
 \hline
  $2.604E^{-04}$ & $1.011\times 10^{-01}$ \hspace{0.1cm} GeV & $H \rightarrow s, S$\\
   \hline
  $2.342E^{-04}$ & $9.092\times 10^{-02}$\hspace{0.1cm} GeV & $H \rightarrow \sim o2, \sim o3 $\\  
   \hline
  $1.549E^{-04}$ & $6.013\times 10^{-02}$\hspace{0.1cm} GeV & $H \rightarrow \sim o4, \sim o4 $\\ 
   \hline
  $1.665E^{-05}$ & $6.464\times 10^{-03}$\hspace{0.1cm} GeV & $H \rightarrow \sim o1, \sim o1$\\ 
   \hline
  $1.282E^{-05}$ & $4.975\times 10^{-03}$\hspace{0.1cm} GeV & $H_{3}\rightarrow \sim o1, \sim o4$\\ 
   \hline
  $4.017E^{-06}$ & $1.559\times 10^{-03}$\hspace{0.1cm} GeV & $H \rightarrow \sim o3, \sim o3$\\
   \hline
  $2.235E^{-06}$ & $8.677\times 10^{-04}$\hspace{0.1cm} GeV & $H_{3}\rightarrow G, G$\\ 
   \hline
  $1.270E^{-07}$ & $4.929\times 10^{-05}$\hspace{0.1cm} GeV & $H_{3}\rightarrow h, h$\\
  \hline  
  $1.411E^{-08}$ & $5.475\times 10^{-06}$\hspace{0.1cm} GeV & $H_{3}\rightarrow Z,Z$\\ 
   \hline
  $5.939E^{-09}$ & $2.305\times 10^{-06}$\hspace{0.1cm} GeV & $H_{3}\rightarrow c, C$\\
   \hline
  $5.774E^{-09}$ & $2.241\times 10^{-06}$\hspace{0.1cm} GeV & $H_{3}\rightarrow ne,Ne$\\  
   \hline
  $2.326E^{-09}$ & $9.028\times 10^{-07}$\hspace{0.1cm} GeV & $H_{3}\rightarrow A, A$\\ 
   \hline
  $1.728E^{-09}$ & $6.706\times 10^{-07}$\hspace{0.1cm} GeV & $H_{3}\rightarrow eL, EL$\\
  \hline 
\end{tabular}
\end{center}
\caption{Branching Ratio, Partial Width and Decay channel of Inert Higgs $ H $ of total width $ 3.88 \times10^{+02} $ .}
\end{table}

\section {Conclusion}

We have proposed a fLipped $SU(5)$ model from $ F $ Theory based on $ A_{4} $ Flavour Symmetry that can interpret the lightest stable dark matter neutralino, its mass and its annhilation cross section via various decay channels. This symmetry $SU(5) \times  U(1)_{\chi}$ breaks down to the SM
gauge group when a $10_{-2} + 10_{2}$ pair of $SU(5) \times  U(1)_{\chi}$ Higgs multiplets acquires Vacuum expectation values. In $SU(5) \times  U(1)_{\chi}$ model, the down type colour triplets of these Higgs representations pair up with the triplets in $ 5+\bar{5} $ Higgs multiplets, and receive large divergent quantum corrections which leds to the suppression of dimension$-$five baryon violating operators. Additionally, the presence of these extra matter fields in the effective model, including an extra pair of right-handed singlets with electric charges ±1, not only contributes to the anomalous magnetic moment of the muon $ g_{\mu} -2 $ but also offers exciting opportunities to explore novel phenomena and signatures that could be probed in ongoing and future experiments. These unique features make the flipped $SU(5) \times  U(1)_{\chi}$ model a compelling candidate for unveiling new physics beyond the Standard Model and advancing our understanding of the fundamental forces in the universe. We also predicted the lifetime of the proton along various decay channels $p \rightarrow e^{+} + K^{0}$, $p \rightarrow e^{+} + \pi^{0}$, $p \rightarrow \mu^{+} + \pi^{0}$, $p \rightarrow \mu^{+} + K^{0}$, for different values of $ tan \beta $, whose values are within current
experimental constraints as determined by the  Super $K$ and Hyper $K$ bounds. It is deduced that the triplets mass is bounded at $ m_{T} \geq 10^{11}$ GeV, for $M_{G} = 10^{16}$ GeV. The different asymptotic values of the lifetime is being imparted by the presence of color Higgs triplets. 

\section{Acknowledgement}
GG would like to thank would like to thank University Grants Commission RUSA, MHRD, Government of India for financial support to carry out this work.

\appendix
\section{$SU(5)\times U(1)$ Symmetry}

In our pursuit of understanding the flipped $SU(5)\times U(1)$ model within a generic F$-$theory framework, we adopt the spectral cover approach and introduce fluxes along $U(1)$ factors to elucidate the geometric properties of matter curves and the massless particle spectrum associated with them. Through this comprehensive analysis, we successfully identify the presence of three generations of chiral matter fields and ascertain the requisite Higgs representations necessary for breaking the symmetry, thereby establishing a robust foundation for further investigations into the model's low-energy implications and phenomenological predictions.  Within each family, the chiral matter fields form a comprehensive $16$ spinorial representation of $SO(10)$, amenable to the insightful $SU(5)\times U(1)_{\chi}$ decomposition.
\begin{equation}
16 = 10_{-1} + \bar{5_{3}} + 1_{-5}
\end{equation}
 Consequently, the Standard Model representations find their embedding as follows:
\begin{equation}
10_{-1} \Longrightarrow F_{i} =  \left(  Q_{i}, d_{i}^{c}, \nu_{i}^{c} \right) 
\end{equation}
\begin{equation}
\bar{5}_{+3} \Longrightarrow \bar{f}_{i} =  \left(  u_{i}^{c}, l_{i} \right) 
\end{equation}
\begin{equation}
1_{-5} \Longrightarrow l_{i}^{c} =  e_{i}^{c}
\end{equation}

The spontaneous breaking of the flipped $SU(5)$ symmetry unfolds through the utilization of a pair of accommodated Higgs fields.

\begin{equation}
H=10_{-1}=\left(Q_{H},d_{H}^{c},\nu_{H}^{c}\right) 
\end{equation}

\begin{equation}
\bar{H} = \bar{10}_{+1}= \left(\bar{Q}_{H},\bar{d}_{H}^{c},\bar{\nu}_{H}^{c}\right) 
\end{equation}
The representation of the MSSM Higgs doublets as fiveplets can be traced back to their origin within the $SO(10)$ group's 10$-$dimensional representation.

\begin{equation}
h = 5_{+2} = \left( D_{h},h_{d}\right), \bar{h} = \bar{5}_{-2}= \left( D_{h},h_{u}\right)
\end{equation}
This $U(1)_{\chi}$ charge assignment not only distinguishes between the Higgs $\bar{5_{-2}}$ fields and the matter anti$-$fiveplets $\bar{5_{-3}}$ in the flipped model, but it also plays a crucial role in generating fermion masses through $SU(5)\times U(1)_{\chi}$ invariant couplings. The generation of fermion masses is attributed to the interaction terms involving $SU(5) \times U(1)_{\chi}$ invariant couplings.
\begin{equation}
W = \lambda_{d}10_{-1}.10_{-1}.5^{h}_{2} + \lambda_{u}10_{-1}.\bar{5}_{3}.\bar{5}^{\bar{h}}_{-2} + 
\lambda_{d}Q.d^{c}.{h}_{d} + \lambda_{u}\left( Q u^{c}h_{u} + l\nu^{c}h_{u}\right) + \lambda_{l}e^{c}lh_{d}
\end{equation}
Notably, the GUT$-$scale predictions of the flipped model establish a distinct relationship between up-quark and neutrino Dirac mass matrices, characterized by $m_{t} = m_{\nu_{D}}$. However, contrary to the standard $SU(5)$ model, the flipped model introduces a disparity in the origins of down quark and lepton mass matrices due to their dependence on separate Yukawa couplings. Shifting focus to the Higgs sector, the acquisition of significant vacuum expectation values (VEVs) by H and $\bar{H}$ of the order MGUT induces the breaking of $SU(5) \times U(1)_{\chi}$ to the Standard Model gauge group, while simultaneously conferring substantial masses upon the color triplets, as evidenced by the ensuing mass terms.
\begin{equation}
HHh + \bar{H}\bar{H}\bar{h} \Rightarrow <\nu_{H}^{c}>d_{H}^{c}D + <\bar{\nu_{H}^{c}}>\bar{d_{H}^{c}}\bar{D}
\end{equation}
Furthermore, an additional higher-order term responsible for imparting Majorana masses to right-handed neutrinos takes the following form:
\begin{equation}
\textit{W} = \frac{1}{M_{s}}\bar{10_{\bar{H}}}\bar{10_{\bar{H}}}10_{-1}10_{-1} \Rightarrow \frac{1}{M_{s}} <\bar{\nu^{c}_{H}}^{2}\nu_{i}^{c}\nu_{j}^{c}
\end{equation}

\end{document}